\documentstyle[prb,aps,multicol,epsfig]{revtex}
\tighten
\newcommand{\beq}{\begin{equation}}
\newcommand{\eeq}{\end{equation}}
\newcommand{\bea}{\begin{eqnarray}}
\newcommand{\eea}{\end{eqnarray}}   

\begin{document}
\draft
\title{Complexation of a polyelectrolyte with oppositely charged
spherical macroions: Giant inversion of charge.}
\author{Toan T. Nguyen and Boris I. Shklovskii}
\address{Department of Physics, University of Minnesota, 116 Church
St. Southeast, Minneapolis, Minnesota 55455} \maketitle
\begin{abstract}
  Complexation of a long flexible polyelectrolyte (PE) molecule with
  oppositely charged spherical particles such as colloids, micelles,
  or globular proteins in a salty water solution is studied.  PE binds
  spheres winding around them, while spheres repel each other and form
  almost periodic necklace.  If the total charge of PE in the solution
  is larger than the total charge of spheres, repulsive correlations
  of PE turns on a sphere lead to inversion of the net charge of each
  sphere. In the opposite case when the total charge of spheres is
  larger, we predict another correlation effect: spheres bind to the PE in
  such a great number that they invert the charge of the PE.  The inverted
  charge by absolute value can be larger than the bare charge of PE
  even when screening by monovalent salt is weak.  At larger
  concentrations of monovalent salt, the inverted charge can reach
  giant proportions.  Near the isoelectric point where total charges
  of spheres and PE are equal, necklaces condense into macroscopic
  bundles.  Our theory is in qualitative agreement with recent
  experiments on micelles-PE systems.
\end{abstract}


\begin{multicols}{2}
\section{Introduction}

Electrostatic interactions play an important role in aqueous solutions
of biological and synthetic polyelectrolytes (PE).  The complexation
of a long flexible polyelectrolyte with oppositely charged spherical
particles such as micelles, \cite{Dubin} globular proteins
\cite{Kabanov} or colloids \cite{Sivan} is a generic electrostatic
problem of the polymer physics.  A long PE binds oppositely charged
spheres winding around each of them (Fig. 1).  Without losing the
generality, we assume that the PE is negative and spheres are positive.

If the charge of a sphere is not completely compensated by the winding
PE, the net charge of the sphere is still positive, the neighbouring
spheres repel each other and form on the PE an almost periodic necklace
(Fig. 1). The same picture is true when the winding PE inverts the net
charge of each sphere making it negative. We call this nontrivial
phenomenon a sphere charge inversion (SCI).  SCI is known to happen in
the most famous biological example of PE-spheres complexation. In the
chromatin, the negative double-helix DNA molecule winds around a
positive histone octamer to form a complex known as the nucleosome
bead.  Nucleosome beads are connected by DNA linkers in the so-called
beads-on-a-string structure.  When linkers are cut enzymatically each
nucleosome bead is found to have a negative net charge.

The counterintuitive phenomenon of SCI has attracted a lot of
attention of theorists.  However, all theoretical and numerical
studies of SCI, have been done for the complexation a single sphere
with a PE molecule
~\cite{Linse,Pincus,Bruinsma,Joanny2,Sens,Stoll,Nguyen3}.

In this paper, we propose the first theory of the SCI in the necklacelike
complex of the PE with many spheres. Our theory accounts for the
interaction between different spheres.  We argue that in this case, as
in the case of a single sphere~\cite{Nguyen3}, SCI happens due to
repulsive correlations of different PE turns on the surface of
spheres.

For many spheres, however, not only the net charge of a sphere should
be found, but simultaneously the number of spheres attached to PE
molecules is to be calculated. Therefore, the second and even more
challenging problem is to determine the sign of the whole complex of
PE with many spheres. Is it positive or negative at given number
concentrations of PE, $n_p$, and spheres, $n_s$, in solution?

The standard Debye-H\"{u}ckel and Poisson-Boltzmann theories of
screening of PE by monovalent counterions leave the net charge of PE
always negative.  These theories, however, do not work for screening
by strongly charged spheres which, as we mentioned above, form a
correlated sequence, reminding a necklace or a one-dimensional Wigner
crystal.  One can call it a Wigner liquid, because the long range
order in many practical situations is destroyed.
\begin{figure}
\epsfxsize=9.0cm \centerline{\epsfbox{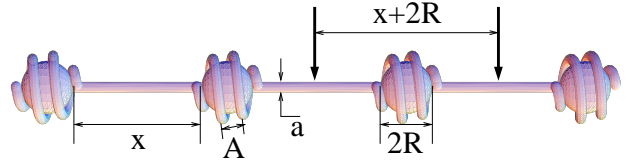}}
\caption{The beads-on-a-string complex of a negative PE molecule 
  and many positive spheres.  On the surface of each sphere, due to
  the Coulomb repulsion, neighbouring PE turns lie parallel to each
  other. Locally, they resemble an one-dimensional Wigner crystal with
  the lattice constant $A$.  At a larger scale, charged spheres repel
  each other and form another one-dimensional Wigner crystal along the
  PE with lattice constant $x+2R$. A Wigner-Seitz cell of this crystal
  is shown by the thick arrows.}
\end{figure}
Wigner-crystal-like correlations between multivalent counterions are
known to lead to charge inversion of rigid macroions
\cite{Nguyen3,Shklov99,Nguyen1,Nguyen2,Joanny,Rubin,Joanny2}.  This
happens because when a multivalent ion approaches an already
neutralized macroion, it repels other counterions, creates for itself
a correlation hole or an image of opposite charge which attracts it to
the surface.  In the necklace shown above, the PE segment wound around
each sphere interacts exclusively with this sphere and plays the role
of the correlation hole or a Wigner-Seitz cell.  Therefore, again,
Wigner-crystal-like correlations come into play and lead to an
additional attraction of the spheres to the PE.  Indeed, when a new
sphere approaches a neutralized necklace, it pushes other spheres
away, unwinds a segment of PE from them and winds this segment around
itself.  This segment is the sphere's correlation hole or, in other
words, its image in the PE.  We are dealing with the correlation
physics because the image appears only in response to the new sphere.
We show below that, at large $n_s$ and small $n_p$, this correlation
attraction leads to PE charge inversion (PECI). PECI was observed in a
micelle-PE system~\cite{Dubin}.

Following Ref.\onlinecite{Shklov99,Nguyen1,Nguyen2} for a quantitative
characteristic of charge inversion we introduce the charge inversion
ratio of the PE ${\cal P} = -Q^* /Q$, where $Q=L\eta$ is the negative
bare charge of PE ($L$ and $\eta$ are, respectively, the contour
length and the linear charge density of a PE molecule) and $Q^*$ is
its positive net charge together with all adsorbed spheres.
Optimization of the free energy of a complex with respect of the
number of bound spheres per PE molecule, $N$, shows that, even for a
large Debye-H\"{u}ckel screening radius $r_D$ of the solution, the
optimal $N=N_0$ is so large that
\beq
{\cal P} =\left(\frac{q}{R\eta\alpha}\right)^{1/4} \gg 1~~.
\label{CIintro2}
\eeq
Here $R$ and
$q$ is the radius and charge of a sphere and
$\alpha$ is a dimensionless logarithmic function
of $qR/\eta r_D^2$ (see Sec. III).
We assume everywhere in this paper that $q/\eta R\gg 1$, so that
more than one turn of PE winds around the sphere to
neutralize it. 

PECI also grows with stronger screening (smaller $r_D$). 
For $r_D$ in the range $A\ll r_D \ll R(q/R\eta)^{1/2}$, we show that
\beq
{\cal P}=\sqrt{\frac{q}{r_D\eta\beta}} \gg
         \left(\frac{q}{R\eta}\right)^{1/4} \gg 1~~.
\label{CIintro}
\eeq
Here $\beta$
is a dimensionless function of $q/R\eta$ and $R/r_D$ (see Sec. V).
A PECI given by Eq. (\ref{CIintro2}) and Eq. (\ref{CIintro}) 
can be called giant.

At a large sphere concentration $n_s$, when the PE number concentration,
$n_p$, grows
and reaches $n_s/N_0$, the pool of free spheres gets exhausted and 
each PE molecule can not get the optimal number $N_0$ of them any more. 
Then PECI becomes weaker and disappears linearly at
the isoelectric point $n_{pi} = qn_s/|Q|$, where the total charge
of all the spheres compensates the total charge of all PE molecules.
When the concentration $n_p$ continues to grow beyond 
$n_{pi}$ practically all the spheres remain bound
to PE and the net charge $Q^*$ is negative and grows by absolute value.
Variation of $Q^*/Q$ with $n_p$ is shown on Fig. 2 by the solid line.
\begin{figure}
\epsfxsize=7.0cm \centerline{\epsfbox{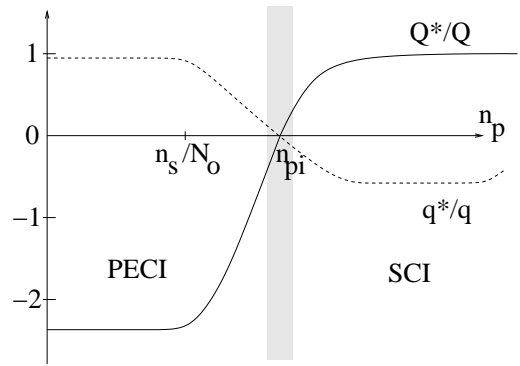}}
\caption{Schematic plot of $q^*/q$ and $Q^*/Q$ as functions
of the PE concentration, $n_p$, at a fixed and large sphere 
concentration $n_s$ and at a not very large $r_D$.
The shaded stripe
corresponds to the region around the isoelectric point
$n_p=n_{pi}$ where necklaces condense into macroscopic bundles.}
\end{figure}
Simultaneously with these variations of $Q^*/Q$,
the net charge $q^*$ of a sphere changes, too.
At $n_p < n_s/N_0$ it is positive and close to $q$.
At $n_p > n_s/N_0$, the net charge $q^*$ starts to decrease linearly
with $n_p - n_{pi}$. 
At the isoelectric point $n_p = n_{pi}$, the charge $q^*$ crosses zero and  
simultaneously, the linker length $x$ vanishes.
At $n_p > n_{pi}$, the charge $q^*$ becomes negative and SCI appears. 
We show that the charge inversion ratio of a sphere,
${\cal S} = - q^*/q$, 
grows with $n_p - n_{pi}$ until it reaches the value
corresponding to a single sphere bound to infinite PE~\cite{Nguyen3},
which is roughly equal to the inverse number of turns 
necessary for PE to neutralize a sphere.
The behavior of $q^*/q$ as function of 
$n_p$ is shown by dashed line in Fig. 2. It is clear from Fig. 2 that 
SCI happens at $n_p > n_{pi}$ and PECI at $n_p < n_{pi}$.

Thus, we arrive at the conclusion that both at $n_p > n_{pi}$
and $n_P < n_{pi}$,
a beads-on-a-string structure can spontaneously self-assemble from
a PE and oppositely charged spheres without any non-Coulomb forces.
The latter structure resembles the 10nm fiber structure of
the chromatin.

Experimental observation of SCI is possible when 
spheres with winding PE are cut out from the complex. 
Then their charge can be measured by electrophoresis. 

Consequences of PECI are more pronounced. PECI leads
to reentrant condensation of necklaces into macroscopic bundles.
Indeed, near the isoelectric point $n_p = n_{pi}$ each complex 
is almost neutral and short range attractive forces
between Wigner-crystal-like complexes~\cite{Rouzina96}
lead to their condensation and coaservation.
Away from the isoelectric point each necklace complex is charged and
their long range repulsive interactions
prevent their condensation. One can watch how condensation
begins and ends changing one of concentrations.
For example, if we keep the spheres concentration $n_s$ large and fixed 
and  start from $n_p\gg n_{pi}$, the complexes
are negative and repel each other. Then with decreasing $n_p$
the condensation happens in the vicinity of the isoelectric point (the
shaded region in Fig. 2). If we
continue decreasing $n_p$, PECI begins and the complexes
become positive.  
When their positive charge, $Q^*$, becomes large enough, the coaservate
dissolves.
An important prediction of such theory~\cite{Rouzina99}
is that the electrophoretic
mobility changes sign in the coaservation range.
We estimate the width of the range of $n_p$ around
$n_{pi}$ where coaservation occurs. This width increases with
decreasing $r_D$ and at $r_D \ll A$:
\beq
\frac{\delta n_p}{n_{pi}}=
        \left(\frac{R\eta}{q}\right)^2\frac{R}{r_D} ~~~.
\label{cowidth}
\eeq

The narrow range of coaservation followed by resolubilization
was observed in the micelles-PE system~\cite{Dubin} as a function of the
charge of micelles. The electrophoretic mobility of complexes was 
indeed found to change
sign within the interval of the micelle
charge in which coaservation happens. The width of the coaservation
region was also observed to increase with decreasing $r_D$ in 
qualitative agreement with Eq. (\ref{cowidth}).

To illustrate the physical picture discussed above
we carry out Monte-Carlo simulation of the complexation
of a negative PE with two positively charged spheres. The system is
in a salt free solution. The simulated spheres have charge
70$e$ uniformly distributed over their surface and radius
3.5$l_B$, where $l_B=7.2\AA$ is the Bjerrum length at the room temperature.
The PE is modeled as a chain of free jointed hard spherical beads
with radius $0.2l_B$ and charge $-e$. The bond length is kept fixed and
equal $l_B$. The Monte-Carlo algorithm is described in our previous
paper (Ref. \onlinecite{Nguyen3}).

\begin{figure}
\epsfxsize=9.0cm \centerline{\epsfbox{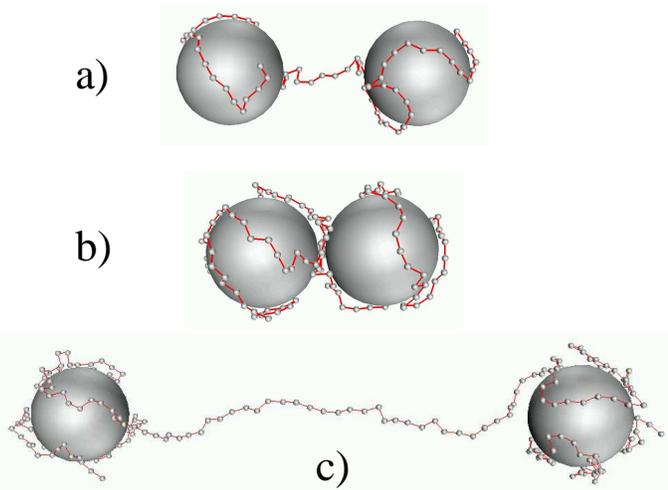}}
\caption{Snapshots of three complexes of a negative PE with two
positively
charged spheres. The numbers of monomers of the PE in the cases (a),
(b) and (c) are 70, 140 and 210 respectively. All the spheres 
have charge 70$e$. The total charges of the complexes are 70$e$,
0 and $-70e$ respectively.
}
\label{Q70snap}
\end{figure}

The snapshots of three such complexes are shown in Fig. \ref{Q70snap}.
In the first simulation, the PE molecule has 70 monomers. 
This complex illustrates the 
regime where the spheres are in abundance
($n_p < n_s/N_0$).
In the second simulation, the PE molecule has 140 monomers so that the
complex is neutral and 
 illustrates the PE-spheres complexes near the isoelectric point
$n_p=n_{pi}$. In the last simulation, the PE molecule has 210 monomers. This 
complex illustrates the regime where
there are not enough spheres to neutralize the PE ($n_p > n_{pi}$). 
PECI is clearly observed in the first simulation. One PE with charge
-70$e$ complexes with the two spheres with
charge 70$e$ each, making a giant 100\% PECI. Around the isoelectric point,
the distance between the surfaces of the two spheres is practically zero 
(less than the length of one PE bond $l_B$). 
Many such PE-spheres complexes condense
into a large bundle around the isoelectric point. 
Far beyond the isoelectric
point, the PE-spheres complex is stretched again. 
SCI is observed with around 85 PE monomers bound to each sphere
($\sim$20\% SCI). 

Although a perfect solenoid conformation
of PE is not observed in Fig. 3,
one can clearly see that PE segments of different turns
stay away from each other and locally resemble a one dimensional
Wigner crystal which helps to lower
the energy of the system. Globally, thermally excited soft bending modes
with characteristic length $R$ melt the solenoid
into a compromised
``tennis ball" conformation~\cite{Nguyen3}. The difference in energy
between a ``tennis ball" and a solenoid, however, is small compare to the 
interaction between the spheres and the PE. This explains
the agreement between the observed physical features
 of the simulated finite temperature systems and
those predicted by our zero temperature theory.

Here, we also would like to mention recent
Monte-Carlo simulations~\cite{Marie} of complexation of a PE molecule
of given length with many oppositely charged spheres.
Results of this work are in qualitative
agreement with Fig. 2. However, we cannot
compare them with our theory quantitatively
because in these simulations the parameter $q/R\eta$ is not large.

It should also be noted that the behaviour of charge inversion for
PE-spheres complexes described above is qualitatively 
similar to that of lipid-DNA complexes studied in Refs. 
\onlinecite{Radler,Bruinsma3} where one sees both kinds 
of charge inversion as one moves away from the isoelectric point.

This paper is organized as follows: In Sec. II, 
we study the free energy of the system and derive
equations for the equilibrium values of $x$ and $N$.
In sections III and IV, these equations are solved for
the weak screening case where the screening radius $r_D$ is larger
than the necklace period $x+2R$. In Sec. V, we discuss condensation
and resolubilization near the isoelectric point.
Section VI is devoted to the strong screening case,
$r_D \ll x+2R$. We show that in this case PECI is much stronger. 
In Sec. VII, we derive the charge inversion ratio for a stiff (rod-like) PE
and compare it to the result for an intrinsically flexible PE obtained in
previous sections. It is shown that at weak screening, PECI is stronger
for flexible PE while at strong screening, PECI is stronger for rigid
PE. This means that if one stretches the PE by external force,
some spheres leave the PE in the weak screening case and condense
on the PE in the strong screening case. 
Finally, in the conclusion, we discuss several important
assumptions used in this work.

\section{Optimization of the complex structure. Weak screening case}

Let us start by writing down the free energy of the complex of a PE with
length $L$ and charge density $\eta$ winding consequently 
around $N$ oppositely charged
spheres of charge $q$ and radius $R$ (see Fig. 1).
First, we assume the complex is in a low salt solution so
that the screening radius $r_D$ is larger than the distance between 
two neighbouring
spheres $x+2R$. We call this situation the weak screening case.
Taking into account that the length of the PE segment that winds around each
sphere is $(L/N-x)$, we have
\beq
F(N,x)=\frac{Q^{*\,2}}{DN(x+2R)}\ln {\cal N}+Nf(x)~~,
\label{ftotal}
\eeq
where
\bea
f(x)&=&\frac{q^{*\,2}}{2RD}-\frac{2q^*\eta}{D}\ln\frac{x+2R}{2R}+
                \nonumber \\
&&      +(x+2R)\frac{\eta^2}{D}\ln \frac{x+2R}{2R} -
        (x+4R)\frac{\eta^2}{D}\ln \frac{x+4R}{4R} \nonumber \\
&&      +(L/N-x)\frac{\eta^2}{D}\ln\frac{A}{a} 
      +x\frac{\eta^2}{D}\ln\frac{x}{2a} .
\label{feach}
\eea
Here $D$ is the dielectric constant of water.
At a length scale greater than its period $x+2R$,
the complex is a uniform rod of length $N(x+2R)$ and
charge density $Q^*/N(x+2R)$. The first term in Eq. (\ref{ftotal})
is the self-energy of this necklace (the macroscopic self-energy). 
The logarithmic divergence of this energy is cut off at small distances
by $x+2R$ and at large distances by the length
\beq
{\cal N}(x+2R)=\min \left\{r_D, N(x+2R)\right\}~~,
\eeq
where $N(x+2R)$ is the rod length.
In the second term of Eq. (\ref{ftotal}), $f(x)$ accounts for the 
total energy of one period of the necklace.
It is calculated as the energy of a Wigner-Seitz cell consisting
of a sphere with two PE tails of length $x/2$. The first terms
in Eq. (\ref{feach}) accounts for
 the self-energy of the adsorbed sphere with net charge
 $q^*$ at the
PE. The second term accounts for the interaction of the sphere with the tails,
the third and fourth terms account for the interaction
between the tails. The fifth and sixth terms are, respectively,
the self-energies 
of the PE wound around the macroion (which is screened at distance $A$
between turns)
and of the two straight tails with length $x/2$. It should be noted
that writing down
the second of Eq. (\ref{ftotal}) as $Nf(x)$ we have neglected the difference
between the end spheres with those in the middle of the PE. This
is justified for a reasonably large value of $N$.
It should also be noted that we neglected the entropy of the
PE monomers in the tails and at the spheres surface. This
is justified because the charge of the sphere is large
and Coulomb energy is much larger than the thermal energy of PE.

As we will see later, when $n_p$ is away from 
the isoelectric point $n_{pi}$, the linker 
length $x$ is much larger than $R$. This helps to simplify
Eqs. (\ref{ftotal}) and (\ref{feach}).  Approximating
$A\simeq R^2/(L/N-x)$ and keeping only terms
of the highest order in the large parameter $x/R$, one can rewrite 
these equations as
\bea
F(N,x)&=&\frac{\delta^2}{x}N\ln {\cal N}+Nf(x) 
\label{ftotal2} 
\\
f(x)&=&\frac{(\delta+x)^2}{2R}
        -2(\delta+x)\ln\frac{x}{R} \nonumber \\
&&      -(L/N-x)\ln\frac{(L/N-x)a}{R^2}+x\ln\frac{x}{a} ~~~,
\label{feach2}
\eea
where we introduce the PE length needed to neutralize
one sphere ${\cal L}=q/\eta$ and 
\beq 
\delta={\cal L}-L/N=Q^*/N\eta~~,
\eeq
so that $q^*=\eta(\delta+x)$. From now on, we also write the energy
in units of $\eta^2/D$ (hence, the energy has 
dimensionality of length).

At a given $N$, the optimal distance $x$ can be calculated by
minimizing the free energy $F(N,x)$ with respect to $x$. 
This gives, to the leading terms,
\beq
\frac{\partial F}{\partial x} =
        -\frac{\delta^2}{x^2}\ln {\cal N}+\frac{\delta+x}{R} 
        -\ln \frac{x}{R}+\ln\frac{L/N-x}{R}=0 ~.
\label{xminimize}
\eeq

The physical meaning of each term in Eq. (\ref{xminimize}) is quite clear.
When one brings a unit length of the PE 
from the sphere surface to their tails,
thereby increasing $x$, the four terms 
of Eq. (\ref{xminimize}) are, respectively, the lowering in the system's
macroscopic energy (with increasing $x$), the potential energy cost due 
to the attraction of the PE to the sphere,
the potential energy gained due to the
repulsion of two PE tails of each sphere
and finally the cost in the correlation energy at the surface
of the sphere. This last term - the correlation energy term - 
needs further clarification.
If the PE turns around a sphere were randomly oriented,
its self-energy per unit length would be $\ln(R/a)$. In reality,
due to strong lateral repulsion between
different PE turns, they lie parallel to each other and
locally resemble a one-dimension Wigner crystal. In this ordered
state, the self-energy per unit length of the PE turn is screened
at distance $A$ instead of $R$. This gives the energy $\ln(A/a)$ per unit
length of the PE. The lowering in the self-energy of
the PE segment wound around a sphere (with length ($L/N-x$))
in the ordered state as compared to the randomly oriented
state is equal to 
$(L/N-x)[\ln(R/a)-\ln(A/a)] = (L/N-x)\ln(R/A) \simeq (L/N-x)\ln((L/N-x)/R)$
and is
called the correlation energy. 
The fourth term of Eq. (\ref{xminimize}) is 
its derivative with respect of $x$.
A more detail discussion of this
correlation effect can be found in Ref. \onlinecite{Nguyen3}.

In principle, one can solve Eq. (\ref{xminimize}) numerically for $x$
as a function of $N$ and other parameters of the system
$L/R$ and ${\cal L}/R$. After that, one can substitute $x(N)$ back into
Eq. (\ref{ftotal2}) and find the optimal value $N_0$ from the equation:
\beq
\left.\frac{{\rm d}F(N,x(N))}{{\rm d}N}\right|_{N=N_0} = \mu_s~~,
\label{mus}
\eeq
where $\mu_s$ is the chemical potential of spheres in the bulk solution.

If the PE concentration is small ($n_p < n_s/ N_0$)
then
$N_0$ and $x(N_0)$ define the configuration of the complex.
However, in the case $n_p > n_s/N_0$ there are less
than $N_0$ spheres for each PE. In this case, $N=n_s/n_p$
(with an exponentially small correction)
and $x(N)$ defines the configuration of the
complex.

Let us now study asymptotic limits 
in which Eq. (\ref{xminimize}) can be solved
analytically providing clear physical picture of our system.

\section{A single polyelectrolyte molecule in concentrated 
solution of spheres. Weak screening case}

In this section we consider the case where the PE concentration is
small, $n_p < n_s/N_0$, so that the optimization of $F(N,x(N))$
with respect to $N$ is needed to get the optimal 
configuration of the complex.
Here and everywhere in this paper we assume
the bulk sphere concentration is high enough so that one can 
approximate $\mu_s={\cal L}^2/2R$ (the self-energy
of a bare sphere) neglecting the entropic part
of the chemical potential. Equation (\ref{mus}) can be
rewritten as
\bea
\frac{{\cal L}^2}{2R}&=&\frac{{\rm d}F}{{\rm d}N}=
        \frac{\delta^2}{x}\left(1+\frac{2L}{N\delta}
        -\frac{Nx^\prime}{x}\right)\ln {\cal N}+
        \nonumber \\
&&      +\frac{(\delta+x)^2}{2R}\left(1+\frac{2L}{N(\delta+x)}+
        \frac{2Nx^\prime}{\delta+x}\right)-
        \nonumber \\
&& \mbox{\hspace{-1cm}} -(2{\cal L}+x+Nx^\prime)\ln\frac{x}{2R}+
        (x+Nx^\prime)\ln\frac{L/N-x}{R}
\label{Nminimize}
\eea
where $x^\prime={\rm d}x/{\rm d}N$.

To solve Eq. (\ref{xminimize}) for $x$, 
we assume that
${\cal L} \gg R\ln {\cal N}$
or, in other words, the screening length is smaller
than an exponentially large length, $r_D \ll x \exp({\cal L}/R)$.
As we see below, in this case $\delta \gg x$ and the last two 
terms in Eq. (\ref{xminimize}) can be neglected. This gives
\beq
x = \delta^{1/2} (R\ln {\cal N})^{1/2}~.
\label{x1}
\eeq
Substituting Eq. (\ref{x1}) into Eq. (\ref{Nminimize}) and keeping
only the highest order terms one obtains the equation
\beq
(\delta R\ln {\cal N})^{1/2}(2\delta+3L/N)-(L/N)^2/2=0~,
\eeq
which has consistent solution only if $\delta \gg L/N$. In this case,
\beq
\delta \sim \frac{L}{N} \left(\frac{L/N}{R\ln {\cal N}}\right)^{1/3}~,
\label{delta1}
\eeq
and the solution for $N_0=L/(x+2R)$ is
\beq
N_0=\frac{L}{\delta^{3/4} (R\ln {\cal N})^{1/4}} 
\simeq \frac{L}{{\cal L}^{3/4}(R\ln {\cal N})^{1/4}}~~.
\label{n01}
\eeq
The corresponding charge inversion ratio is
\beq
{\cal P}=-\frac{Q^*}{Q}=\frac{N_0\delta}{L}=
\left(\frac{\cal L}{R\ln {\cal N}}\right)^{1/4}
\gg 1~.
\label{CIweak1}
\eeq
From Eq. (\ref{x1}) and (\ref{delta1}), it is easy to see that
the relative order of all the lengths in the system is
\bea
\delta  &\simeq& {\cal L} \gg L/N={\cal L}^{3/4}(R\ln {\cal N})^{1/4} 
\nonumber \\
        &\gg& x={\cal L}^{1/2}(R\ln {\cal N})^{1/2}\gg R\ln {\cal N}\gg R~.
\label{scale1}
\eea
This order is consistent with the assumptions we started with.

As we saw above,
the two last logarithmic terms in Eq. (\ref{Nminimize}) are negligible.
Therefore, the main driving force behind PECI 
is the gain in the self-energy of a sphere when PE winds
around it reducing its net charge. It is the
difference between the left hand side and the second term 
of Eq. (\ref{Nminimize}). 
In other words, the sum of the self-energies $q^{*\,2}/2RD$ 
decreases when PE distributes 
itself over larger number of spheres. This correlation effect 
overcomes the macroscopic energy cost of overcharging the
PE (the first term on the right hand side of Eq. (\ref{Nminimize})).
Therefore, PECI can be well obtained in the approximation
where PE charge is smeared on the surface of 
spheres\cite{Pincus,Bruinsma}.

Let us explain why we still call PECI calculated here a correlation
effect.
As we saw above the reason for this PECI is that each sphere is bound to
several
turns of a negatively charged PE. These turns can be considered as a
correlation hole in the sense that this is the part of PE, which
interacts almost exclusively with the given sphere (other spheres are at
much larger distance $x \gg R$).
The segments of PE wound around each sphere have the same length,
$L/N-x \sim L/N \gg x$. Therefore, similarly to 
Ref.~\onlinecite{Nguyen3,Shklov99,Nguyen1,Nguyen2} we are
dealing with Wigner-crystal-like correlations and the wound
segment can be considered as a Wigner-Seitz cell of the bare sphere.
The gain in the sphere self-energy mentioned above is nothing but the usual
binding energy per sphere of a Wigner crystal: the interaction of a sphere
with its Wigner-Seitz cell.
 
Note that because most of the PE length is wound around
the spheres, the periodicity of positions of spheres covered by PE
solenoids in the real space
(see Fig. 1) is less important than in the case of a rigid
PE (see Sec. VII) or other cases of charge inversion of rigid macroions
by multivalent counterions~\cite{Shklov99,Nguyen1,Nguyen2}.
Linkers between different pairs of neighbouring spheres may differ 
in their length without a substantial change in the two major contributions 
to the free energy discussed above
(the sphere self-energy gain and the macroscopic charging energy).
The thermal motion can even melt the Wigner crystal of spheres in the
real space while the length of the wound segment remains unchanged.
Therefore PECI is much more robust than the Wigner crystal
in the real space.

\section{High concentration of polyelectrolyte. 
Weak screening case.}

In this section, we deal with the case when $n_p > n_s/N_0$ and there is
shortage of spheres, each PE cannot get the optimal number, 
$N_0$, of spheres found in previous section. In this case,
the number of spheres per PE is fixed: $N=n_s/n_p$.
Therefore
\beq
\frac{Q^*}{Q}=\frac{Q-n_s q/n_p}{Q}=1-\frac{n_{pi}}{n_p}
        =\frac{\delta}{\cal L}~,
\label{PECI2}
\eeq
so that PECI becomes weaker and linearly decreases with
$n_p-n_{pi}$ as $n_p$ grows.
When $n_p$ increases beyond the isoelectric point $n_{pi}=n_s q/Q$, 
the total charge of the complex $Q^*$ is negative. 
The ratio $Q^*/Q$ increases linearly from zero and eventually saturates
at unity as $n_p$ increases further. The behavior of $Q^*/Q$
as function of $n_p$ is plotted  by the solid curve in Fig. 2.

Let us now discuss the behavior of the net charge of
the sphere $q^*=\eta(\delta+x)$ as $n_p$ increases. To do so,
one has to solve Eq. (\ref{xminimize}) and 
find the distance $x$ by which the spheres are separated along the PE
(we stress again that we are interested in the complex far enough from the
isoelectric point, so that $x\gg R$ and Eq. (\ref{xminimize}) is valid.)

As $n_p$ increases beyond $n_s/N_0$, the last two logarithmic terms in
Eq. (\ref{xminimize}) are still negligible compare to the second term.
Therefore, $x$ is given by Eq. (\ref{x1})
(it should be noted that, here, 
$\delta={\cal L}(1-n_p/n_{pi})$ is a given length).
Correspondingly, $q^*$ decreases linearly with $\delta$.

As $n_p$ moves closer to the isoelectric point,
the net charge $q^*=\eta(\delta+x)$ decreases. When
\beq
\delta < \delta_c=R\ln\frac{L/N-x}{R} \simeq R\ln({\cal L}/R)~~,
\label{border1}
\eeq
(here, we replace $\ln((L/N-x)/R)$ by $\ln({\cal L}/R)$  because
near the isoelectric point, $\delta, x \ll {\cal L} \sim L/N$)
the fourth and the first terms of Eq. (\ref{xminimize}) start to 
dominate over the second and third terms. This gives
\beq
x\simeq |\delta| \sqrt{\frac{\ln{\cal N}}{\ln ({\cal L}/R)}}~.
\label{xweak}
\eeq

Of course, Eqs. (\ref{x1}) and (\ref{xweak}) match each other at
$\delta=\delta_c$.
To continue, let us consider the two important limiting cases
$\ln{\cal N} \ll \ln({\cal L}/R)$
and 
$\ln{\cal N} \gg \ln({\cal L}/R)$.

\subsection{The case $\ln{\cal N} \ll \ln({\cal L}/R)$}
In this case, $x \ll |\delta|$ and, therefore, the charge
 of a sphere $\eta(\delta+x)$, decreases to zero
and becomes negative as $n_p$ passes through $n_{pi}$ (see Fig. 2).
At $n_p > n_{pi}$, this SCI is
driven by the fourth term of Eq. (\ref{xminimize}): the correlation
energy of PE segment at the surface of the spheres.
The charge inversion ratio 
${\cal S}=|\delta+x|/{\cal L}\simeq |1-n_p/n_{pi}|$ (see Fig. 2).

As $n_p$ increases further, the charge of the sphere $\delta+x$
grows and the second and the fourth terms of Eq. (\ref{xminimize}) 
become the dominant ones. This gives
\beq
q^*/\eta=\delta+x=-R\ln\frac{L/N-x}{R}\simeq -R\ln\frac{\cal L}{R}~,
\eeq
so that the charge inversion ratio reaches its maximal possible value
(see Fig. 2) which is equal to that for the complexation of a single
sphere and a polyelectrolyte~\cite{Nguyen3}:
$$
{\cal S}=\frac{-q^*}{q} \simeq \frac{R}{\cal L}\ln\frac{\cal L}{R}~.
$$
Therefore, roughly speaking, it is inversely proportional to 
the number of turns of PE around the sphere.

As $n_p$ continues to increase, $x$ increases and the third term
of Eq. (\ref{xminimize}) becomes important making the sphere
charge less negative. When $x > {\cal L}$, the second and
third terms of Eq. (\ref{xminimize}) dominate. This gives
\beq
q^*/\eta=\delta+x=R\ln\frac{x}{R}\simeq R\ln\frac{L}{NR}~,
\eeq
so that the the net charge of the sphere changes sign from negative 
back to positive
(not shown in Fig. 2).
However, the condition of low salt solution, $x < r_D$, 
assumed in the derivation of Eq. (\ref{xminimize}),
makes this re-entrant inversion of charge unrealistic.
In practical situation, $r_D < {\cal L}$
so that the necklace remains in the SCI range. A detail consideration
of the strong screening case $r_D < x$ is presented in the next section.

\subsection{The case $\ln{\cal N} \gg \ln({\cal L}/R)$}
In this case, Eq. (\ref{xweak}) gives $x \gg |\delta|$ 
and the charge of a sphere $\eta(\delta+x)$ touches zero but
stays positive as $n_p$ passes through the isoelectric point
despite the fact that the total charge of
the complex $Q^*=N\delta$  changes from positive to negative.
This is because for a long PE, the macroscopic energy is very
large and the complex is under a strong
stress to increase $x$ in order to reduce this macroscopic energy. 
This decreases the amount of PE that
can wind around each sphere making the sphere positive. 

As $n_p$ increases beyond the isoelectric point, $|\delta|$,
$x$ and the net charge $q^*=\delta+x$ of each sphere
increase as well. 
Eventually $q^* \simeq q$, 
the PE unwinds from all of its spheres and
becomes a straight rod to which $N$ spheres are attached to. 
Substituting $x=L/N$ into Eq. (\ref{xminimize}) and neglecting the last
two terms of this equation, one can estimate
the value $L/N$ at which PE unwinds from the spheres:
$L/N \simeq {\cal L}(1+\sqrt{{\cal L}/R\ln{\cal N}})$.
As $n_p$ increases further, $q^*/q$ saturates at unity. The
behavior of $q^*/q$ as the function of $n_p$ is depicted by
the dashed line in Fig. 4.
\begin{figure}
\epsfxsize=7.0cm \centerline{\epsfbox{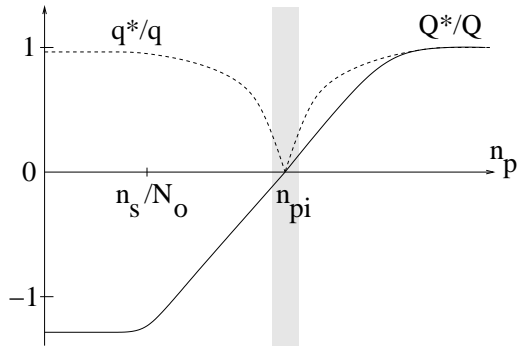}}
\caption{Schematic plot of $q^*/q$ and $Q^*/Q$ as functions
of PE concentration at a fixed and large sphere concentration $n_s$
for the case B, Sec. IV. There is
no SCI in this case. The shaded stripe shows the range of $n_p$
around $n_{pi}$ where condensation of PE molecules happens.}
\end{figure}
We would like to emphasize that the inequality
$R\ln{\cal N} \gg \ln({\cal L}/R)$ may require unreasonably large screening
radius $r_D$, so that the behavior presented in Fig. 2 for case A
of this section is more generic.

\section{Condensation of PE-spheres complexes near 
  the isoelectric point.}  Now, let us discuss properties of the
system near the isoelectric point.  Exactly at the isoelectric point
$n_p=n_{pi}$, the spheres-PE complex is neutral, $Q^*=q^*=\delta=x=0$
and $L/N=\cal L$. From Eq. (\ref{ftotal2}) one gets the energy of one
complex as $L\ln(A/a)$. It is the self energy of the PE $L\ln(R/a)$
(the PE is straight up to distance $R$) plus the correlation energy
$-L\ln(R/A)$ gained by arranging PE turns into one-dimensional Wigner
crystal at the sphere surface (see the discussion after Eq.
(\ref{xminimize})).  A consequence of this interpretation for the
energy is that at the isoelectric point PE molecules condense onto
each other forming a macroscopic neutral bundle. This is because the
density of PE in the region where the spheres touch each other (the
region bounded by broken line in Fig. 5) is doubled. Thus, the
distance between PE segments, $A_t$, is halved, $A_t=A/2$, which
results in a gain in the correlation energy.  Simple geometrical
calculation shows that this region has the area $AR$. Therefore, the
PE in this region has total length $R$.  The correlation energy gain
per unit volume is
$$
\Delta E_{corr}\sim -\frac{n_pL}{\cal L}R\ln\frac{A}{A_t}
        \sim -\frac{n_pL}{\cal L}R~~.
$$
\begin{figure}
\epsfxsize=8.0cm \centerline{\epsfbox{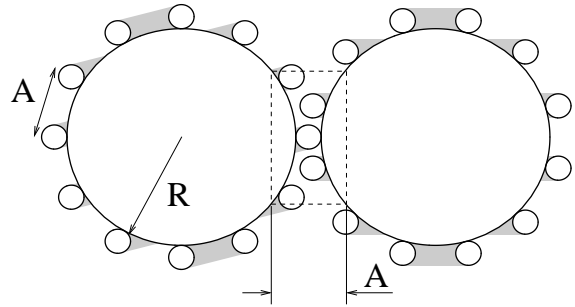}}
\caption{Cross section through the centers of two touching spheres 
  with worm-like (gray) PE wound around them.  At the place where two
  spheres touch each other (the region bounded by the broken line) the
  density of PE and the background surface charge doubles which in
  turn leads to a gain in the correlation energy of PE segments.  Near
  the isoelectric point, this gain is responsible for the condensation
  of spheres-PE complexes, forming a large neutral bundle of PE-
  spheres complexes.}
\label{2spheresfig}
\end{figure}
Because of this finite gain in the correlation energy, there is a
finite range of $n_p$ around $n_{pi}$ that the PE molecules are still
in a condensed state. Let us try to find the width of this region.

To find the boundary of the condensation region on the left side of
the isoelectric point ($n_p < n_{pi}$), one needs to compare the total
energy of the system in the condensed and dissolved states.  Here, the
condensed state contains a macroscopic neutral bundle of PE-spheres
complexes and $(n_s-n_pL/\cal L)$ leftover spheres per unit volume.
The bundle is neutral because charging a macroscopic body costs a lot
of energy.  The dissolved state is a solution of $n_p$ isolated
PE-spheres complexes per unit volume, each PE adsorbing $n_s/n_p$
spheres.  At the condensation concentration $n_p=n_{pl}$, we have to
balance the correlation energy gain $\Delta E_{corr}$ with the loss in
the self energy of $(n_s-n_pL/\cal L)$ left-over spheres when they
change from almost-neutralized spheres at the PE molecules to bare
spheres in solution.  Therefore the equation for the condensation
point $n_p=n_{pl}$ is
\beq
\frac{n_{pl}L}{\cal L}R=
\left(n_s-\frac{n_{pl}L}{\cal L}\right) \frac{{\cal L}^2}{2R}
\label{collapsel}
\eeq
or
\beq
1-\frac{n_{pl}}{n_{pi}} \simeq \frac{R^2}{{\cal L}^2}~~.
\eeq

On the right side of the isoelectric point ($n_p > n_{pi}$),
the condensed state is a macroscopic neutral bundle of PE and spheres
and ($n_p-n_{pi}$) leftover bare PE molecules. $\Delta E_{corr}$ needs
to be balanced with the lost in the self-energy of PE molecules
when they change from almost-neutralized state to bare state in solution.
This gives for the resolubilization concentration $n_{pr}$:
\beq
\frac{n_{pi}L}{\cal L}R=(n_{pr}-n_{pi})L\ln\frac{r_D}{a}
\label{collapser}
\eeq
or
\beq
\frac{n_{pr}}{n_{pi}}-1 \simeq \frac{R}{{\cal L}\ln(r_D/a)}~~.
\eeq
Finally, the total width of the region, $\Delta n_p=n_{pr}-n_{pl}$, 
around $n_{pi}$ where condensation occurs is
\beq
\frac{\Delta n_p}{n_{pi}}=\frac{R^2}{{\cal L}^2}+
        \frac{R}{{\cal L}\ln(r_D/a)} \simeq \frac{R}{{\cal L}\ln(r_D/a)}~~.
\eeq
Comparing this with Eq. (\ref{border1}), we see that the width of the
condensation region is small, well within the region where
the correlation energy (the fourth term in Eq. (\ref{xminimize}))
is important in determining conformation of the system. Therefore,
in this range $\Delta E_{corr}$ indeed dominates all other
energies in Eq. (\ref{ftotal2}) as we assumed.

\section{Strong screening by monovalent salt.}

Until now, we assumed the salt concentration is small enough so that
the screening radius $r_D$ is larger than the distance between
neighbouring spheres, $x$. In the case of higher salt concentration
when $r_D \ll x$, our theory needs some modifications. First, the
macroscopic energy term (the first term in Eq. (\ref{ftotal})) has to
be replaced by the sum of repulsion energies of neighbouring spheres.
When $R\ll r_D \ll x$, it still has the form of interaction of two
point-like charges:
\beq
F(N,x)=N\frac{q^{*\,2}}{x+2R}e^{-(x+2R)/r_D}+Nf(x)~~.
\label{fscreen}
\eeq
At the same time, all the logarithmic factors in Eq. (\ref{feach}) for
$f(x)$ are cut off at $r_D$ instead of $x$.
Correspondingly, Eq. (\ref{xminimize}) (which is the result of the
minimization of $F(N,x)$ with respect to $x$ at a given $N$) should
be replaced by
\bea
\frac{\partial F}{\partial x}&=&-\frac{(\delta+x)^2}{x r_D}e^{-x/r_D}
        +\frac{\delta+x}{R}-\ln\frac{r_D}{R} \nonumber \\
&&      ~~+\ln\frac{L/N-x}{R}=0~.
\label{xminscreen1}
\eea

Let us concentrate on the PECI regime when $n_p < n_s/N_0$.
 In this case, the last two logarithmic terms in Eq.
(\ref{xminscreen1}) can be neglected. This gives
\beq
x=r_D\ln\frac{(\delta+x) R}{r_D x}\simeq r_D \ln\frac{{\cal L}R}{r_D^2}~~.
\label{xscreen1}
\eeq
Thus, the condition $r_D \ll x$ is equivalent to $r_D \ll \sqrt{{\cal
    L}R}$. In this case $r_D \ll x \ll \delta \sim {\cal L}$.  One can
see that $x$ only weakly depends on the number $N$ of spheres attached
to the PE. This is because the macroscopic self-energy of the complex
which forces the PE to unwind from the sphere is strongly screened and
diminished at distances beyond $r_D \ll x$.

Substituting Eq. (\ref{xscreen1}) back into Eq. (\ref{fscreen}) and
optimizing $F(N,x(N))$ with respect to $N$, we have, to the leading
term
\beq
\frac{(L/N)^2}{2R}=\frac{r_D(\delta+x+2L/N)}{R}+\frac{{\cal L}x}{R}~~,
\label{Nscreen1}
\eeq
so that
\beq
\frac{L}{N}=\sqrt{{\cal L}x}=\sqrt{{\cal L}r_D\ln\frac{{\cal L}R}{r_D^2}}
\eeq
and the charge inversion ratio is
\beq
{\cal P}=\frac{N\delta}{L}=\sqrt{\frac{{\cal L}/r_D}{\ln({\cal L}R/r_D^2)}}~~.
\label{CIscreen1}
\eeq

Comparing these results with those of Sec. III 
we see that due to screening, the spheres are closer to each other
($x\propto r_D$ instead of ${\cal L}^{1/2}R^{1/2}$) and a smaller
length of the PE is wound around a sphere.  In other words, the
positive net charge of each sphere is larger ($L/N\propto {\cal
  L}^{1/2}r_D^{1/2}$ instead of ${\cal L}^{3/4}R^{1/4}$).  Therefore,
more spheres are attached to the PE, making charge inversion much
stronger (${\cal P} \propto {\cal L}^{1/2}$ instead of ${\cal
  L}^{1/4}$).  At the same time, when $r_D$ increases to about
$\sqrt{{\cal L}R}$, $x \sim r_D$, $L/N\simeq {\cal L}^{3/4} R^{1/4}$,
${\cal P}\simeq ({\cal L}/R)^{1/4}$ and we come back to the weak
screening case.

It should be stressed that, for optimization with respect to $x$, the
gain in a sphere's self-energy when PE winds around the sphere (the
second term on the right hand side of Eq. (\ref{xminscreen1})) is
balanced with the repulsion from its neighbouring spheres (the first
term). However, when determining $N$ and $\cal P$ from Eq.
(\ref{Nscreen1}), the repulsion between the spheres described by the
first term on the right hand side, which is of the order ${\cal
  L}r_D/R$, is negligible compared to the second term ${\cal L}x/R$.
This term originates from the fact that when one brings a sphere from
solution to the PE, hence gains the self-energy $(L/N)^2/2R$, the PE
unwinds from other spheres in order to prepare the linker $x$ for this
new sphere.

Let us now consider even smaller screening radius $r_D \ll R$.
In this case, one has to modify
all the energy terms of Eq. (\ref{fscreen}).
The self-energy of each sphere becomes $q^{*\,2}r_D/2R^2$ instead of
$q^{*\,2}/2R$ and
the interaction between neighbouring spheres is
$$
\frac{(q^*r_D^2/R^2)^2}{x}e^{-x/r_D}~~.
$$
As a result, the minimization with respect to $x$ gives
\beq
x=r_D\ln\frac{(\delta+x)r_D^2}{xR^2}\simeq r_D\ln\frac{{\cal L}r_D}{R^2}~~.
\label{xscreen2}
\eeq
Now, an equation
similar to Eq. (\ref{Nscreen1}) gives
\beq
\frac{L}{N}=\sqrt{{\cal L}x}=\sqrt{{\cal L}r_D\ln\frac{{\cal L}r_D}{R^2}}~~
\label{Nscreen2}
\eeq
and 
\beq
{\cal P}=\frac{N\delta}{L}=\sqrt{\frac{{\cal L}/r_D}{\ln({\cal L}r_D/R^2)}}~~,
\label{CIscreen2}
\eeq
so that the charge inversion is indeed stronger in this case and
increases even faster than Eq. (\ref{CIscreen1}) with decreasing
$r_D$.  Equations (\ref{CIscreen1}) and (\ref{CIscreen2}) match each
other when $r_D \sim R$. Equation (\ref{CIintro}) is their
combination.

\begin{figure}
\epsfxsize=8.0cm \centerline{\epsfbox{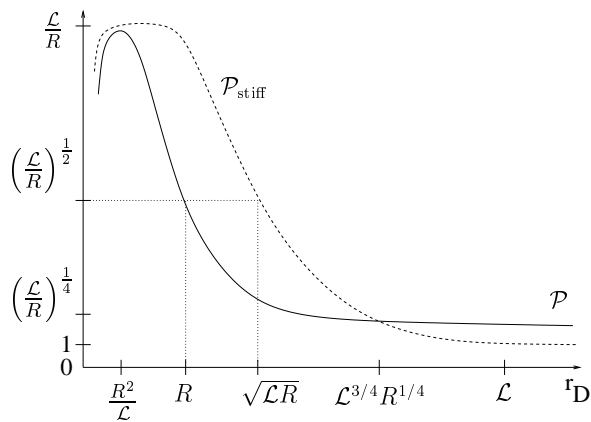}}
\caption{Schematic plot of the charge inversion ratios $\cal P$ 
(the solid line) and 
$\cal P_{\rm stiff}$ (the dashed line) as function
of screening length $r_D$. 
${\cal P}_{\rm stiff} > {\cal P}$ at 
$r_D \ll {\cal L}^{3/4}R^{1/4}$. $\cal P$ saturates
when $r_D\sim R^2/{\cal L}$ while ${\cal P}_{\rm stiff}$ saturates
when $r_D \sim R$. For the definition of ${\cal P}_{\rm stiff}$, see
Sec. VII.
}
\label{PvsrD}
\end{figure}

When the screening length becomes smaller than $R^2/{\cal L}$,
the logarithmic factor $\ln({\cal L}r_D/R^2)$ should be replaced by unity
and Eqs. (\ref{xscreen2}) and (\ref{Nscreen2}) give $L/N\sim R$ and $x\ll R$.
This means that PE is a straight rod with
the bare spheres closely packed on it. The number of spheres attached to
the PE reaches
its maximal possible value $N=N_{\rm max}=L/R$, 
and so does the charge inversion ratio 
${\cal P}={\cal P}_{\rm max}={\cal L}/R$.
The behavior of $\cal P$ as a function of the screening length $r_D$
is shown by the solid line in Fig. \ref{PvsrD}. 
It should be noted that
at very small value of the screening length, when the energy of interaction
between a sphere and the PE 
is less than $k_BT$, 
the spheres detach from 
the PE and $\cal P$ rapidly decreases to zero.

Until now we have concentrated on the effect of strong screening on 
the optimal
configuration of the PE-spheres complex at small concentration $n_p < n_{s}/N_0$, when
spheres are in abundance. Now we want to discuss the role of screening in
the opposite case, $n_p > n_{s}/N_0$. 
In this case, the number of spheres per PE
$N=n_s/n_p$ is fixed and Eq. (\ref{PECI2}) remains valid
(screening does not affect $Q^*$, because in any
case all spheres are adsorbed by PE.) 
Qualitatively, Fig. 2 remains valid in this case.
The length $x$ is now given by the first
equality of Eq. (\ref{xscreen1}) for $r_D>R$ and by Eq. (\ref{xscreen2})
for $r_D<R$ (the second equalities in these equations is not valid
because $\delta$ is a given length). Therefore, $x$ decreases as
$\delta$ decreases.

In the case $r_D > R$, when $\delta$ decreases below $r_D^2/R$,
Eq. (\ref{xscreen1}) gives $x \leq r_D$ and we
come back to the weak screening case described in Sec. IV, case A.
All discussion about SCI and condensation of complexes in Sec. IV, case A remains
valid in this case.

On the other hand, in the case $A < r_D < R$, our theory needs some
correction. The value of $\delta$ below which the correlation
energy between PE turns at the surface of a sphere is important is given by
\beq
\delta < \frac{R^2}{r_D}\ln\frac{L/N-x}{R} 
        \simeq \frac{R^2}{r_D}\ln({\cal L}/R)~~,
\label{border2}
\eeq
instead of Eq. (\ref{border1}). This is because, in this case, 
the second term in Eq. (\ref{xminimize}) is $(\delta+x)r_D/R^2$
instead of $(\delta+x)/R$.

At the same time, SCI effect at $n_p > n_{pi}$ is strongly enhanced in
a way similar to the case of one sphere~\cite{Nguyen3}. This is
because the charging energy cost for SCI is strongly suppressed at
small $r_D$ while the short-range correlation energy between PE turns
responsible for SCI remains unaffected. At small screening length,
${\cal S}$ can be larger than unity.

Strong screening ($A \ll r_D \ll R$) also affects the range of $n_p$
where PE molecules form neutral macroscopic bundle.  When $r_D \ll R$,
the self-energy cost ${\cal L}^2/2R$ in the right hand side of Eq.
(\ref{collapsel}) has to be replaced by ${\cal L}^2r_D/R^2$ per sphere
while the short-range correlation energy on the left hand side remains
unaffected. This increases the width of this region to
\beq
\frac{\Delta n_p}{n_{pi}} 
        \sim \frac{R^3}{{\cal L}^2r_D}+\frac{R}{{\cal L}\ln(r_D/a)}~~.
\label{cowidthscreen}
\eeq
This width continue to grow with decreasing $r_D$. When $r_D \sim
R^2/{\cal L} \sim A$, $R^3/{\cal L}^2r_D \sim R/{\cal L}$ and the
width more than doubles. When $r_D < A$, one can neglect the second
term of Eq. (\ref{cowidthscreen}) and arrive at Eq. (\ref{cowidth}).
This equation predicts a strong growth of $\Delta n_p/n_{pi}$ with
decreasing $r_D$, in qualitative agreement with experimental results
of Ref.~\onlinecite{Dubin}.  It should be noted again that, as one see
from comparison with Eq. (\ref{border2}), this coaservation range is
well within the region of $\delta$ where the correlation energy
between PE turns is the dominant energy term.

Finally, when $n_p \gg n_{pi}$, there is very small number of spheres per PE
such that the length of the PE linker between them is larger than
the optimal $x$ given by Eq. (\ref{xscreen1}) and (\ref{xscreen2}),
the linker is no longer straight and each sphere with PE wound around
it behaves independent from each other. SCI saturates at that given
for the case of one sphere - one PE complexation~\cite{Nguyen3}.

\section{Polyelectrolyte with extremely large persistence length}

In this section we assume that the PE has an extremely large persistence
length such that it cannot wind around a sphere. 
In this case, the PE is a straight
rod to which the spheres are attached to. We are interested in
the PECI regime where the concentration of spheres is large.

For a rod-like PE, $x=L/N$, $Nx^\prime=-L/N=-x$, $\delta+x=\cal L$.
In the case of weak screening, $r_D \gg L/N$,
Eq. (\ref{Nminimize}) can be rewritten as
\beq
{\cal L} \frac{N\delta}{L}\ln {\cal N} \simeq {\cal L}\ln\frac{L/2N}{R}~~.
\label{muhard}
\eeq
The physical meaning of this equation is very simple: the left hand
side is the macroscopic charging energy cost when a sphere is brought
from the bulk solution to the PE. The right hand side is the gain in
the correlation energy of the Wigner crystal of spheres along the PE
which helps to overcome the charging energy cost. This correlation
energy is the interaction of the sphere with two PE tails of length
$L/2N$ which forms a Wigner-Seitz cell.

The charge inversion ratio can be easily calculated
\beq
{\cal P}_{\rm stiff}=\frac{N\delta}{L}=\frac{\ln(L/2NR)}{\ln {\cal N}} 
\simeq \frac{\ln({\cal L}/R)}{\ln {\cal N}}.
\label{CIstiffweak}
\eeq

In the case of strong screening, $r_D \ll L/N$, the macroscopic
charging energy cost should be replaced by the repulsion between
neighbouring spheres. At the same time, the logarithmic term in the
expression for the correlation energy of the Wigner crystal of spheres
along the PE should be cut off at $r_D$ instead of $L/N$. Eq.
(\ref{muhard}) now reads:
\beq
\frac{{\cal L}^2}{r_D}e^{-L/Nr_D}={\cal L}\ln\frac{r_D}{R}~~,
\eeq
which gives
$L/N\simeq r_D\ln({\cal L}/r_D)$,
and the charge inversion ratio
\beq
{\cal P}_{\rm stiff}=
        \frac{N\delta}{L}\simeq\frac{\cal L}{r_D\ln({\cal L}/r_D)} \gg 1~.
\label{CIstiffstrong}
\eeq

Let us now compare these results with those for an intrinsically flexible
PE case studied in Sec. III (weak screening) 
and Sec. V (strong screening). 

At weak screening (Sec. III) $r_D > {\cal L} > R\ln {\cal N}$,
Eq. (\ref{CIweak1}) and Eq. (\ref{CIstiffweak}) give
$$
\frac{{\cal P}_{\rm stiff}}{\cal P}= 
\frac{R\ln({\cal L}/R)}{{\cal L}^{1/4}(R\ln{\cal N})^{3/4}}
\simeq \frac{\ln({\cal L}/R)}
{({\cal L}/R)^{1/4}(\ln{\cal N})^{3/4}} \ll 1~.
$$
The last inequality is due to ${\cal L}/R \gg 1$ and $\ln{\cal N} \gg 1$.

As $r_D$ decreases further so that ${\cal L} > r_D > \sqrt{{\cal L}R}$
we enter the strong screening regime for the rod-like PE but still
stay in the weak screening regime for flexible PE.  Using Eq.
(\ref{CIstiffstrong}) for ${\cal P}_{\rm stiff}$ and Eq.
(\ref{CIweak1}) for ${\cal P}$ , we can easily see that the charge
inversion for the rod-like PE starts to become stronger than charge
inversion for the flexible PE when $r_D \sim {\cal L}^{3/4}R^{1/4}$.

When $r_D$ continues to decrease in the range $r_D < \sqrt{{\cal
L}R}$, we are in the strong screening regime for both types of PE.
Equation (\ref{CIscreen1}) and (\ref{CIscreen2}) show that ${\cal P}$
is of the order of $\sqrt{{\cal L}/r_D}$ which is much smaller than
${\cal P}_{\rm stiff} \simeq {\cal L}/r_D$.

The behavior of the charge inversion ratio as a function of $r_D$ for
the two types of PE is shown in Fig. \ref{PvsrD}.  One can explore a
transition from the flexible PE case to the stiff PE case by applying
an external stretching force to the PE\cite{Grosberg}.  To describe
this phenomenon, one has to add to the free energy (\ref{ftotal2}) of
the complex an additional term $-{\cal F} N(x+2R)$, where ${\cal F}$
is the external force. This new term is linearly proportional to the
length of the spheres-PE complex. It adds a negative term $-{\cal F}$
to the right hand side of Eq. (\ref{xminimize}) and therefore
increases $x$. One then can proceed in exactly the same way as in Sec.
III to find the conformation of the complex.  At weak screening when
$r_D \gg {\cal L}^{3/4}R^{1/4}$ (Sec. III), it is not difficult to
show that $x$ increases linearly with the strength of the external
force when this force is small.  At the same time, $N_0$ decreases
linearly with ${\cal F}$ so that one by one, the spheres leave the PE
as the force increases and PECI becomes weaker.  When ${\cal F} \sim
(\eta^2/D)({\cal L}/R - \ln{\cal N})$ (so that the force helps to
balance the attractive potential of the sphere with the macroscopic
repulsive potential of the complex), one obtains $x \sim L/N$ and the
PE unwinds completely from the spheres and becomes straight.  The
problem of complexation of PE and spheres, then, becomes that of a
stiff PE described at the beginning of the section.  This sequential
release of spheres is similar to the problem of stretching a PE
necklace in poor solvent~\cite{Strech}.

The picture is completely reversed at the strong screening case
when $r_D \ll {\cal L}^{3/4}R^{1/4}$. In this case, as one stretches
the PE, the spheres come to the PE one by one and make
the charge inversion stronger.
The strength of the force at
which PE unwinds completely from the spheres can be
calculated in exactly the same way as in the weak screening case.
In the strong screening case, however, the macroscopic repulsive
potential of the complex is very small so that the external force
has to overcome only the attractive potential of
the sphere in order to unwind the PE molecule. Therefore,
 PE unwinds completely when
${\cal F}\sim \eta^2{\cal L}/RD$ for $r_D > R$ and 
${\cal F}\sim \eta^2{\cal L}r_D/R^2D$ for $r_D < R$.

It would be interesting to verify experimentally that the spheres 
leave the PE at weak screening and condense on the PE at 
strong screening when the PE undergoes an external stretching force.

\section{Conclusion}
In conclusion, we would like to discuss four most important
approximations used in this papers.  Let us start from the use of the
Debye-H\"{u}ckel linear theory to describe screening by monovalent
salt.  It is known that if a PE molecule or a sphere are charged
strongly enough this linear approximation does not work and the
nonlinear condensation of counterions takes place, which leads to a
renormalization of their charge.  For the case of a rod-like PE this
phenomenon is known as the Onsager-Manning
condensation~\cite{Manning}. It happens when the linear charge density
of PE $\eta$ is larger than $\eta_c = Dk_{B}T/e$, where $k_B$ is the
Boltzmann constant and $T$ is the temperature.  Correspondingly,
Debye-H\"{u}ckel theory used above is valid when $\eta < \eta_c$.

The condition for the absence of counterion condensation on the
charged spheres is more complicated and involves the concentration of
monovalent salt as well.  It is known that if a sphere is charged
strongly enough, its counterions condense onto its surface to reduce
its charge to the universal critical value $q_c=DRek_BT\ln(c_s/c)$,
where $e$ is the elementary charge, $c_s \sim q_c/R^3e$ and $c$ are
the counterion concentrations at the sphere surface and in the bulk
respectively~\cite{Gueron}.  The condensation on the spheres can be
neglected if $q={\cal L}\eta$ is less than $q_c$ or
\beq
r_D > R e^{{\cal L}\eta/R\eta_c}~~.
\label{rsmax}
\eeq
At small enough $\eta$ (such that $\eta/\eta_c<R/{\cal L}$), this 
condition reduces to $r_D > R$.

When $r_D <R$, each sphere can be considered as a plane with surface
charge density $q/4\pi R^2$ and it is also known that if $r_D$ is
small enough, a charged plane is linearly screened. Specifically, Eq.
(73) of Ref. \onlinecite{Nguyen2} shows that screening is linear if
\beq
r_D < A e^{\eta_c/\eta}\sim \frac{R^2}{\cal L}e^{\eta_c/\eta}.
\label{rsmin}
\eeq
At small enough $\eta$  (such that $\eta<\eta_c/\ln({\cal L}/R)$),
this condition reduces to $r_D < R$.

Thus our theory has a wide range of applicability. For a PE with 
\beq
\eta < \eta_c R/{\cal L}
\label{finalcond}
\eeq 
it is applicable for any value of the screening length $r_D$.
Remarkably, the same inequality for $\eta$ also guarantees that no
Onsager-Manning condensation occurs on the spheres-PE rod-like complex
with the inverted linear charge density $\eta {\cal P}$, even though
the magnitude of this inverted charge can be much larger than the bare
charge of the PE (${\cal P}\gg 1$).  Indeed, as we already know,
${\cal L}/R$ is the maximal possible value for $\cal P$ (see Fig. 6)
therefore Eq. (\ref{finalcond}) guarantees that $\eta {\cal P}$ is
smaller than the Onsager-Manning critical linear charge density
$\eta_c$.

On the other hand, our theory literally is not applicable to strongly
charged PE such as double helix DNA or too strongly charged spheres.
We do, however, believe that our main results are qualitatively
applicable in this case with properly normalized charges of the
spheres and PE. A more detail study of the condensation problem will
be the subject of a future work.

The second approximation which we want to address here is the
assumption that PE is flexible.  It is valid if the elastic energy of
PE winding around a sphere is smaller than the Coulomb energy of
complexation. This can be true even if the persistence length of the
strongly screened PE is of the order of a sphere diameter or somewhat
larger.  For example, theoretical estimates show that even in the
nucleosome bead, the rigidity of DNA plays a secondary role.  We
should emphasize here that we talk about the intrinsic persistence
length, ignoring the well known Odijk-Skolnick-Fixman electrostatic
enhancement of the persistence length, which can be very substantial
for a free PE molecule in the solution with small concentration of
salt~\cite{OSF}. The reason for this is that at the sphere surface the
charge of PE is screened by the sphere's positive charge and it is the
intrinsic persistence length of PE which determines whether PE can
wrap around a sphere.  All electrostatic interactions are explicitly
taken into account in the correlation picture considered.

If the PE intrinsic persistence length is much larger than $R$, the
ground state of the complex can strongly differ from that in the
flexible case.  In Sec. VI we studied the extreme case when the
persistence length of the PE is infinite and the PE is rod-like. There
is a wide range of intermediate magnitudes of persistence length which
is not studied here. In this range nontrivial star-like configurations
become possible even for one sphere~\cite{Bruinsma2}.  For many
spheres one can imagine different kinds of structures. Near the
isoelectric point, it can be a rod-like complex made of a PE solenoid
densely stuffed by spheres. It can be a similar cylinder where PE is
instead winding around spheres makes several parallel to the cylinder
axis straight strands on the surface of the cylinder, which are
connected by loops at the cylinder edges.  This strands repel each
other and form periodic in polar angle "Wigner crystal".  These and
other possible configurations should be studied in future works.

The third simplification used in this paper was an assumption that the
number and the size of spheres in the solution is fixed. When we try
to compare this theory with the data on the micelles-PE system~\cite
{Dubin} we immediately see that, strictly speaking, this is a
different problem, because in the experiment of Ref. ~\onlinecite
{Dubin} the amount of lipids is fixed, but the number and size of
micelles is determined by equilibrium conditions and may depend, for
example, on the screening radius of the solution. This complication
should be taken into account by a future theory.

The fourth important assumption we made in this paper is that the
concentration of spheres, in the solution, $n_s$, is large enough so
that their entropy can be neglected. This assumption leads, for
example, to the conclusion that in the case $n_p > n_s/N_0$, all the
spheres are consumed by PE. Actually, even in this case, there is a
finite, but exponentially small concentration,$n_{s0}$, of free
spheres in the solution next to PE because the binding energy of a
sphere to PE is finite. In other words, this is the concentration of
the "saturated vapour" of spheres right above the correlated liquid of
spheres on PE.  When total concentration of spheres, $n_s$, is so
small that it is comparable to $n_{s0}$ effect of free spheres becomes
very important. For example, at such small $n_s$ spheres fail to
neutralize PE near isoelectric point.  At the limit of $n_p
\rightarrow 0$ neutralization happens at $n_s = n_{s0}$. Therefore,
the plot of the line $n_s(n_p)$ at which neutralization takes place
deviates from isoelectric line $n_s = n_pL/ \cal L$ at small $n_p$ and
$n_s$ (see dashed line at Fig. 7). Correspondingly, the region of the
condensation of PE-spheres complexes on the plane $(n_s, n_p)$ looks
as the shaded region shown in Fig. 7.
\begin{figure}
\epsfxsize=7.0cm \centerline{\epsfbox{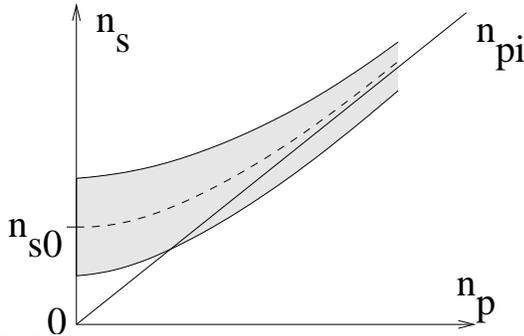}}
\caption{Schematic phase diagram of the condensation of spheres-PE complexes
  in the ($n_p,n_s$)-plane for a very large PE length $L$. Complexes
  condense in the shaded domain. The straight line corresponds to the
  isoelectric point $n_s=n_p L/{\cal L}$. The dash line shows the
  function $n_s(n_p)$ at which an isolated PE-spheres complex is
  neutral.  }
\end{figure}  
Theory of condensation at small $n_p$ is similar to theory of
condensation of DNA by multivalent counterions at small concentration
of DNA \cite{Rouzina99}. Experimental data on re-entrant condensation
of DNA have qualitatively similar shape of the condensation
domain\cite{Raspaud}.

Summarizing our results, we have studied complexation of a negative PE
with positive spherical macroions in salty water.  Under solely the
influence of electrostatic interactions, a PE molecule winding around
individual spheres binds spheres in a beads-on-a-string structure.  At
a large sphere concentration, we found interesting phenomena on both
sides of isoelectric point, at which the total charge of all PE
molecules is exactly compensated by total charge of spheres.  When the
PE concentration is below the isoelectric point, spheres overcharge PE
so that the net charge of PE together with bound spheres is positive
and can be substantially larger than the absolute value of the bare
charge of PE.  When the PE concentration is above the isoelectric
point, the net charge of PE is not inverted, while the net charge of
each sphere together with the PE segment winding around it becomes
inverted and negative. It can be larger than the bare charge of the
sphere if the screening radius of the solution is small enough.
 
In the narrow vicinity of the isoelectric point PE-spheres complexes
condense together in bundles.  We calculated the width of the range of
the PE concentration where the condensation takes place and showed
that it grows very fast with decreasing screening radius of the
solution.

All these phenomena are results of repulsive correlations between PE
turns on the surface of spheres and of the spheres on PE, which can
not be included in any Poisson-Boltzmann-like description of the
mutual screening of PE and spheres. The repulsive correlations between
PE turns on spheres are responsible for charge inversion of individual
spheres (SCI) and the condensation of PE-spheres complexes in bundles.
In the latter case this phenomenon is illustrated by Fig. 5. In the
former case, additional discussion of the physics of this phenomenon
can be found in Ref. \onlinecite{Nguyen3}.  Wigner-crystal-like
correlations of turns mean that each turn is surrounded by a stripe of
positively charge macroion surface, which can be considered as its
positive correlation hole.
 
The most interesting phenomenon of the inversion of charge of PE
(PECI) by a large number of adsorbed spheres is related to the fact
that each sphere is bound to several turns of negatively charged PE.
These turns can be considered as an correlation hole, because this is
the part of PE, which almost exclusively interacts with the given
sphere.  In this sense, once more we are dealing with correlations.
The segments of PE wound around each sphere have the same length and
they constitute most of the PE's total length.  Therefore, these
correlations are similar to Wigner-crystal-like correlations which are
responsible for charge inversion of a rigid macroion
\cite{Nguyen3,Shklov99,Nguyen1,Nguyen2,Rubin}.  This confirms our
point of view that Wigner-crystal-like correlations are the universal
driving force of charge inversion.

Our theory, with minor modifications, can also describe the
complexation of polyelectrolytes with macroions of non-spherical
shape.  An example of such system can be the complexation of two
oppositely charged polyelectrolytes in water solution.  Let us assume
that the negative PE is long and flexible while the positive PE is
shorter, stronger charged and rigid. Then the negative PE molecule
wraps around the positive one and the only change which should be made
in our theory is to replace the expression for self-energy of charged
sphere by the corresponding expression for charged rod. The case when
the positive PE is flexible on the first glance seems to be more
complicated.  However, away from isoelectric point positive PE is
overscreened or underscreened by the wrapping negative one so that it
has a rod-like shape. Therefore, even in this case our theory is valid
with the above-mentioned minor change.

\acknowledgements
We are grateful to A. Yu. Grosberg for helpful discussion and 
reading of the manuscript.
This work was supported by NSF DMR-9985985.

\end{multicols}
\end{document}